\documentclass[12pt,a4paper]{article}
\usepackage{epsfig,graphicx}
\usepackage{amsmath,amsfonts}
\usepackage{amssymb,cite}
\usepackage{t1enc}

\newcommand{\unit}{\leavevmode\hbox{\small1\kern-3.6pt\normalsize1}}

\begin{document}
\vspace*{-3cm}
\begin{flushright}
LC--TH--2003--98 \\
hep-ph/0312140 \\
November 2003
\end{flushright}

\begin{center}
\begin{Large}
{\bf Study of selectron properties in the \\[0.2cm]
$\boldsymbol{\tilde e \, \tilde e \to e^- \tilde \chi_1^0 \, e^- \tilde \chi_2^0}$
decay channel}
\end{Large}

\vspace{0.5cm}
J. A. Aguilar--Saavedra \\[0.2cm]
{\em Departamento de Física and GTFP, \\
Instituto Superior T\'ecnico, P-1049-001 Lisboa, Portugal} \\
\end{center}

\begin{abstract}
We discuss selectron pair production in $e^- e^-$ scattering, in the processes
$e^- e^- \to \tilde e_L \tilde e_L, \tilde e_R \tilde e_R \to e^-
\tilde \chi_1^0 e^- \tilde \chi_2^0 \to e^- \tilde \chi_1^0 e^- \tilde \chi_1^0
f \!\bar f$. This decay channel has in general a smaller branching ratio than
the $\tilde e \tilde e \to e^- \tilde \chi_1^0 e^- \tilde \chi_1^0$ mode, but
has the advantage that the momenta of all final state particles can be
determined without using the selectron masses as input. The reconstruction
of the momenta allows the simultaneous study of: ({\em i\/}) selectron mass
distributions; ({\em ii\/}) selectron spins, via the angular distributions of
the $e^-$ in the selectron rest frames; ({\em iii\/}) selectron masses and
spins, using the $e^-$ energy distributions in the CM frame; ({\em iv\/}) the
selectron ``chirality'', with the analysis of the spin of the produced
$\tilde \chi_2^0$.
\end{abstract}

\section{Introduction}
\label{sec:1}

One of the main motivations for the construction of a linear collider like
TESLA, with centre of mass (CM) energies of $500$ GeV and above, is the precise
determination
of the parameters of supersymmetry (SUSY), if this theory is realised in
nature \cite{tdr}. The analysis of the selectron properties (and in
general the properties of all the sleptons) at a linear collider is of
special interest, since these particles are among the lightest ones in many SUSY
scenarios. Selectron pairs can be copiously produced in $e^+ e^-$ and $e^- e^-$
scattering, being their leading decay mode $\tilde e \to e \tilde \chi_1^0$.
In this note we discuss the determination of the selectron properties
in $\tilde e_L \tilde e_L$ and $\tilde e_R \tilde e_R$ production,
with one of the selectrons decaying to $e^- \tilde \chi_1^0$
and the other decaying via $\tilde e \to e^- \tilde \chi_2^0 \to e^- \tilde
\chi_2^0 f \! \bar f$. This decay mode has generally a smaller cross section
than the $e^- \tilde \chi_1^0 e^- \tilde \chi_1^0$ channel, but has the
advantage that
all the final state momenta can be determined without taking the selectron
masses as input \cite{old}. In our analysis we concentrate on $e^- e^-$
scattering at 500 GeV, but the discussion can be straightforwardly applied
to $e^+ e^-$ collisions, other CM energies and smuon pair production.
This note is organised as follows.
In Section~\ref{sec:2} we discuss the production and subsequent decay of
selectrons in $e^- e^-$ scattering, and briefly explain how these processes
are generated. In Section~\ref{sec:3} we analyse various mass, angular and
energy distributions for the final state
$e^- \tilde \chi_1^0 \, e^- \tilde \chi_1^0 f \! \bar f$.
Our conclusions are presented in Section~\ref{sec:4}.

\section{Generation of the signals}
\label{sec:2}

\begin{figure}[htb]
\begin{center}
\begin{tabular}{ccc}
\mbox{\epsfig{file=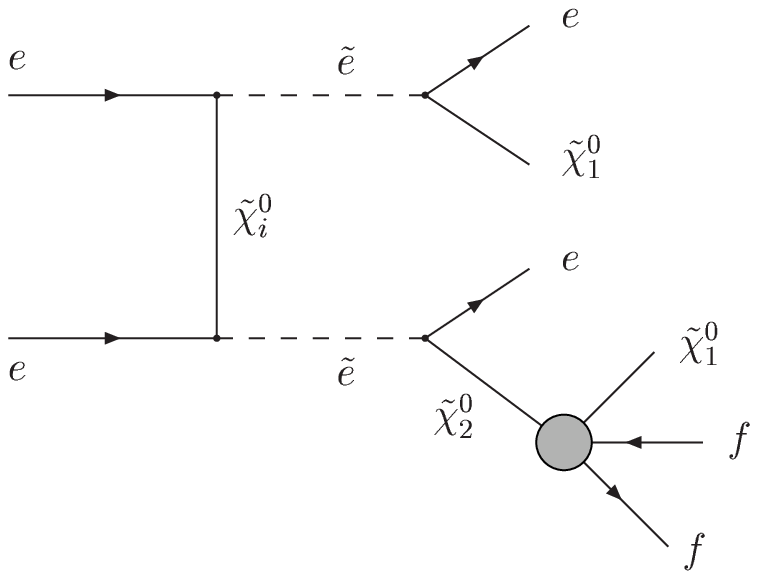,width=6cm,clip=}} & \hspace*{5mm} &
\mbox{\epsfig{file=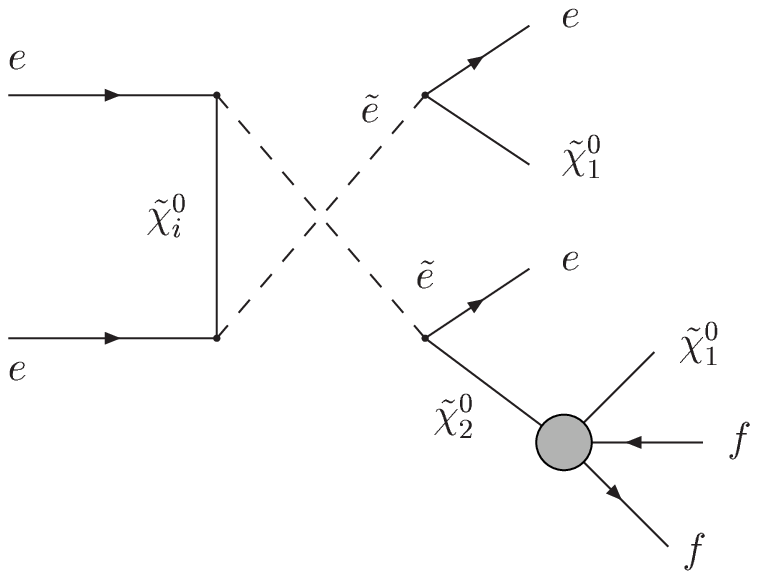,width=6cm,clip=}}
\end{tabular}
\caption{Feynman diagrams for the processes in Eqs.~(\ref{ec:e-e-}).
The 4 neutralinos are exchanged in the $t$ channel, and 
the shaded circles stand for the 8 sub-diagrams mediating the decay of
$\tilde \chi_2^0$, separately shown in Fig.~\ref{fig:X2decay}.}
\label{fig:8legs}
\end{center}
\end{figure}

\begin{figure}[htb]
\begin{center}
\begin{tabular}{ccc}
\mbox{\epsfig{file=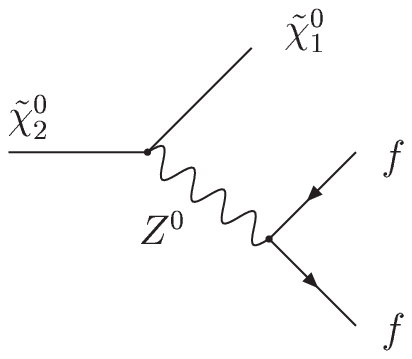,width=3cm,clip=}} & \hspace*{5mm} & 
\mbox{\epsfig{file=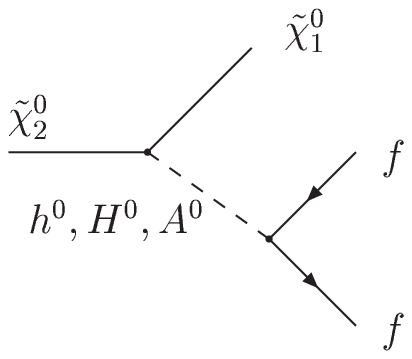,width=3cm,clip=}} \\
(a) & & (b) \\[0.5cm]
\mbox{\epsfig{file=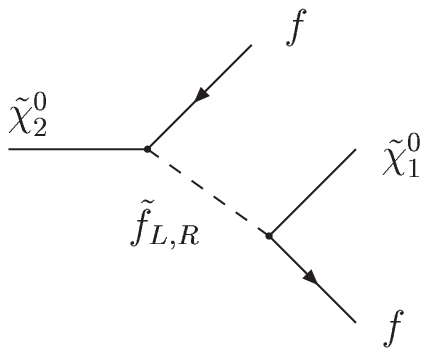,width=3cm,clip=}} & \hspace*{5mm} & 
\mbox{\epsfig{file=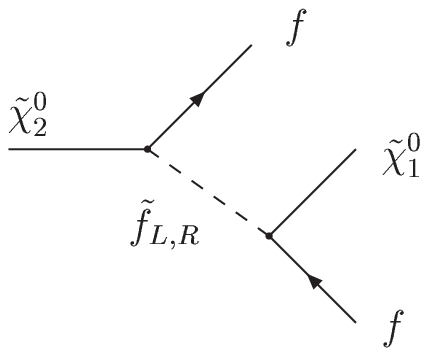,width=3cm,clip=}} \\
(c) & & (d)
\end{tabular}
\caption{Feynman diagrams for the decay of 
$\tilde{\chi}^0_2$, mediated by $Z$ bosons (a), neutral Higgs bosons (b),
and left-and right-handed scalar fermions (c and d).}
\label{fig:X2decay}
\end{center}
\end{figure}

Selectron pairs are produced in $e^- e^-$ collisions through the
diagrams depicted in Fig.~\ref{fig:8legs}, with the four neutralinos $\tilde
\chi_i^0$ exchanged in the $t$ channel.
The Majorana nature of the neutralinos is essential for the
nonvanishing of the transition amplitudes,
as can be clearly seen in this figure. The decay of the $\tilde \chi_2^0$
takes place through the 8 diagrams in Fig.~\ref{fig:X2decay}. In total, there
are 64 diagrams mediating each of the processes
\begin{eqnarray}
& & e^- e^- \to \tilde e_L \tilde e_L \to 
e^- \tilde \chi_1^0 e^- \tilde \chi_2^0 \to
e^- \tilde \chi_1^0 e^- \tilde \chi_1^0 f \!\bar f \,, \nonumber \\
& & e^- e^- \to \tilde e_R \tilde e_R \to 
e^- \tilde \chi_1^0 e^- \tilde \chi_2^0 \to
e^- \tilde \chi_1^0 e^- \tilde \chi_1^0 f \!\bar f \,, \nonumber \\
& & e^- e^- \to \tilde e_R \tilde e_L \to 
e^- \tilde \chi_1^0 e^- \tilde \chi_2^0 \to
e^- \tilde \chi_1^0 e^- \tilde \chi_1^0 f \!\bar f \,.
\label{ec:e-e-}
\end{eqnarray}

The production of mixed $\tilde e_R \tilde e_L$ pairs must be taken into
account as well,
since it constitutes the main background to $\tilde e_L \tilde e_L$ and $\tilde
e_R \tilde e_R$ production in which we are interested.
We only consider final states with $f \! \bar f =
\mu^+ \mu^-, q \bar q$, and in the case of $q \bar q$ we sum
$u \bar u$, $d \bar d$, $s \bar s$, $c \bar c$ and $b \bar b$ production,
without flavour tagging. In the channel
$\tilde \chi_2^0 \to \tilde \chi_1^0 e^+ e^-$, the multiplicity of
electrons in the final state makes it difficult to identify the electron
resulting from the decay of the $\tilde \chi_2^0$. In
$\tilde \chi_2^0 \to \tilde \chi_1^0 \nu \bar \nu$ the presence of
four undetected particles in the final state yields too many unmeasured momenta
for their kinematical determination, and the same happens in
$\tilde \chi_2^0 \to \tilde \chi_1^0 \tau^+ \tau^-$, because each of the $\tau$
leptons decays producing one or two neutrinos that escape detection.

For the generation of the signals we calculate the full matrix elements for the
$2 \to 6$ resonant processes in Eqs.~(\ref{ec:e-e-}), at a CM energy
of 500 GeV and with an integrated luminosity of 100 fb$^{-1}$. In
the calculation we include the effects of initial state radiation (ISR)
and beamstrahlung. We perform a simple simulation of
the detector effects with a Gaussian smearing of the energies, applying also
phase space cuts on transverse momenta $p_T \geq 10$ GeV, pseudorapidities
$|\eta| \leq 2.5$ and ``lego-plot'' separation $\Delta R \geq 0.4$. All the
details concerning the generation of the signals can be
found in Ref.~\cite{old}.

\begin{figure}[htb]
\begin{center}
\epsfig{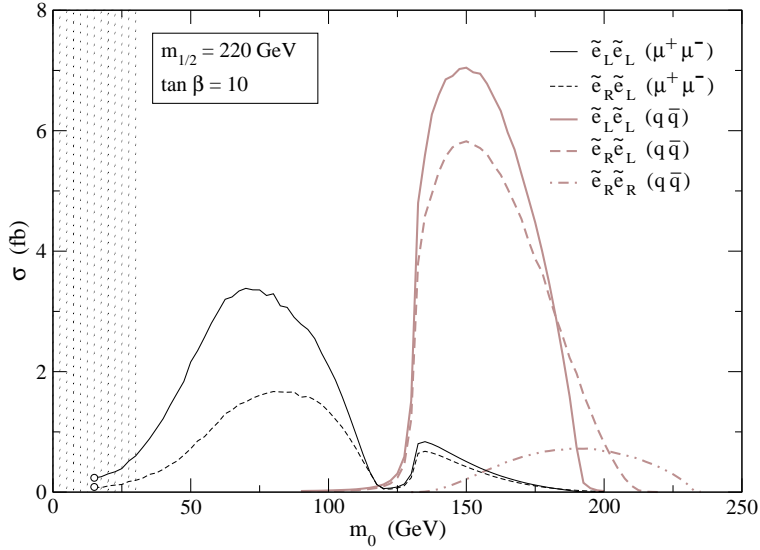}
\caption{Cross sections (in fb) for 
$e^- e^- \to \tilde e \tilde e \to 
e^- \tilde \chi_1^0 e^- \tilde \chi_2^0
\to e^- \tilde \chi_1^0  e^- \tilde \chi_1^0 f \!\bar f$
(with unpolarised beams)
as a function of $m_0$, for $m_{1/2}=220$ GeV, $\tan \beta=10$, $A_0=0$ and
$\mu > 0$.}
\label{fig:m0}
\end{center}
\end{figure}

For our discussion we restrict ourselves to mSUGRA scenarios, requiring them to
be in agreement with present experimental data.
We set $m_{1/2} = 220$ GeV (in order to have a light spectrum so that these
processes are observable with a CM energy of 500 GeV at TESLA), $\tan
\beta = 10$, $\mu > 0$ and for simplicity we choose $A_0=0$ (a more extensive
discussion is presented in Ref.~\cite{old}). The dependence of the studied
signals on the remaining parameter $m_0$ is shown in Fig.~\ref{fig:m0}.

The shaded area on the left of this plot corresponds to values of $m_0$
excluded by the current experimental bounds on $m_{\tilde{e}_R}$.
In this figure we identify two regions of interest for these signals:
\begin{enumerate}
\item For $m_0 \lesssim 120$ GeV  the second neutralino
decays predominantly to charged leptons, $\tilde \chi_2^0 \to \tilde \chi_1^0
l^+ l^-$. The decay amplitudes are dominated by the exchange of on-shell
right-handed sleptons (diagrams (c) and (d) in Fig.~\ref{fig:X2decay}). The
contribution of  diagram (a) with an on-shell
$Z$ boson is less important, due to the smallness of the 
$Z \tilde \chi_2^0 \tilde \chi_1^0$ coupling. 
In this region of the parameter space,
the decays to $\tilde \chi_1^0 q \bar q$ are very suppressed, 
not only because of the small coupling of the $Z$ boson
to $\tilde \chi_1^0$ and $\tilde \chi_2^0$, but also due to
the heavy squark masses, $m_{\tilde q} \gtrsim 400$ GeV.

\item For $m_0 \gtrsim 140$ GeV, the right-handed sleptons
(including $\tilde \tau_R$) are heavier than $\tilde \chi_2^0$, and the
$Z$-exchange diagram in Fig.~\ref{fig:X2decay} dominates, yielding a large
branching ratio for $\tilde \chi_2^0 \to \tilde \chi_1^0 q \bar q$.
\end{enumerate}

Between both regions there is a narrow window $m_0 \sim 130$ GeV where the decay
$\tilde \chi_2^0 \to \tilde \chi_1^0 \tau^+ \tau^-$ completely dominates,
because the $\tilde e_R$ and $\tilde \mu_R$ are heavier than the $\tilde
\chi_2^0$ but the $\tilde \tau_R$ is lighter.
For $m_0 \gtrsim 200$ GeV, the production of $\tilde e_L \tilde e_L$
and $\tilde e_L \tilde e_R$ is not possible with a CM energy of
500 GeV, and only $\tilde e_R \tilde e_R$ pairs are produced.
From inspection of Fig.~\ref{fig:m0} we select two values
$m_0 = 80$ GeV and $m_0 = 160$ GeV to illustrate 
the reconstruction of the selectron
masses for $\mu^+ \mu^-$ and $q \bar q$ final states. The sets of
parameters for these two mSUGRA scenarios are summarised in Table \ref{tab:in}.
The resulting selectron and neutralino masses
and widths, as well as the relevant branching ratios,
are collected in Table \ref{tab:out} for each of these scenarios.

\begin{table}[htb]
\begin{center}
\begin{tabular}{ccccc}
Parameter & ~ & Scenario 1 & ~ & Scenario 2 \\
\hline
$m_{1/2}$ & & 220 & & 220 \\
$m_0$ & & 80 & & 160 \\
$A_0$ & & 0 & & 0 \\
$\tan \beta$ & & 10 & & 10 \\
$\mathrm{sign} \, \mu$ & & $+$ & & $+$ 
\end{tabular}
\caption{Input parameters for the two mSUGRA scenarios to be 
considered in Section
\ref{sec:3}. The values of $m_{1/2}$, $m_0$ and $A_0$ are in GeV.
\label{tab:in}}
\end{center}
\end{table}

\begin{table}[htb]
\begin{center}
\begin{tabular}{cccc}
 & Scenario 1 & ~ & Scenario 2 \\
\hline
$m_{\tilde e_L}$ & 181.0 & & 227.4 \\
$\Gamma_{\tilde e_L}$ & 0.25 & & 0.85 \\
$m_{\tilde e_R}$ & 123.0 & & 185.0 \\
$\Gamma_{\tilde e_R}$ & 0.17 & & 0.58 \\
$m_{\tilde \chi_1^0}$ & 84.0 & & 84.3 \\
$m_{\tilde \chi_2^0}$ & 155.8 & & 156.4 \\
$\Gamma_{\tilde \chi_2^0}$ & 0.023 & & $1.50 \times 10^{-5}$ \\
$m_{\tilde \chi_3^0}$ & 309.4 & & 310.0 \\
$\Gamma_{\tilde \chi_3^0}$ & 1.48 & & 1.43 \\
$m_{\tilde \chi_4^0}$ & 330.4 & & 331.1 \\
$\Gamma_{\tilde \chi_4^0}$ & 2.30 & & 2.01 \\
$\mathrm{Br}(\tilde e_L \to e^- \tilde \chi_1^0)$ & 41.8 \% & & 18.3 \% \\
$\mathrm{Br}(\tilde e_L \to e^- \tilde \chi_2^0)$ & 20.6 \% & & 30.8 \% \\
$\mathrm{Br}(\tilde e_R \to e^- \tilde \chi_1^0)$ & 100 \% & & 99.7 \% \\
$\mathrm{Br}(\tilde e_R \to e^- \tilde \chi_2^0)$ & $\simeq 0$ & & 0.3 \% \\
$\mathrm{Br}(\tilde \chi_2^0 \to \tilde \chi_1^0 \mu^+ \mu^-)$ & 10.3 \% & &
 3.9 \% \\
$\mathrm{Br}(\tilde \chi_2^0 \to \tilde \chi_1^0 q \bar q)$ & $\simeq 0$ & &
 69.2 \% \\
\end{tabular}
\caption{Some relevant quantities in the two mSUGRA scenarios
considered in Section \ref{sec:3}. The masses and widths are in GeV.
\label{tab:out}}
\end{center}
\end{table}

It is worth comparing between the $e^- \tilde \chi_1^0 e^- \tilde \chi_2^0$
decay mode studied here and the leading channel $e^- \tilde \chi_1^0 e^- \tilde
\chi_1^0$. For $\tilde e_L \tilde e_L$ production, in scenario 1 the total
branching ratio (including the decay of the $\tilde \chi_2^0$) of the 
$e^- \tilde \chi_1^0 e^- \tilde \chi_1^0 \mu^+ \mu^-$ signal is 0.88\%, while
for the leading channel it is 17.4\%. On the other hand, in
scenario 2 the total branching ratio of the
$e^- \tilde \chi_1^0 e^- \tilde \chi_1^0 q \bar q$ signal is 3.9\%, slightly
larger than for the
$e^- \tilde \chi_1^0 e^- \tilde \chi_1^0$ channel.
For $\tilde e_R \tilde e_R$ production, in scenario 1 the 
decay $\tilde e_R \to e^- \tilde \chi_2^0$ is not possible because the
$\tilde e_R$ is lighter than the $\tilde \chi_2^0$. In scenario 2,
$\mathrm{Br}(\tilde e_R \to e^- \tilde \chi_2^0) \simeq 0.3$\%
because the $\tilde e_R$
only couples to the small bino component of the second neutralino. The total
branching ratios in each case are collected in Table~\ref{tab:brcomp}. 

\begin{table}
\begin{center}
\begin{tabular}{ccccc}
 & \multicolumn{2}{c}{Scenario 1} & \multicolumn{2}{c}{Scenario 2} \\
Final state & $\tilde e_L \tilde e_L$ & $\tilde e_R \tilde e_R$ 
 & $\tilde e_L \tilde e_L$ & $\tilde e_R \tilde e_R$ \\
\hline
$e^- \tilde \chi_1^0 e^- \tilde \chi_1^0 f \! \bar f$ 
 & 0.88\% & 0 & 3.9\% & 0.27\% \\
$e^- \tilde \chi_1^0 e^- \tilde \chi_1^0$ 
 & 17.4\% & 100\% & 3.3\% & 99\%
\end{tabular}
\end{center}
\caption{Comparison of the branching ratios for the $e^- \tilde \chi_1^0 e^-
\tilde \chi_1^0 f \! \bar f$ and $e^- \tilde \chi_1^0 e^- \tilde \chi_1^0$
signals in scenarios 1 and 2. In scenario 1, $f \! \bar f = \mu^+ \mu^-$, while
in scenario 2 $f \! \bar f = q \bar q$.
\label{tab:brcomp}}
\end{table}

\section{Reconstruction of the final state and kinematical distributions}
\label{sec:3}

To reconstruct the final state momenta we use as input the 4-momenta of the
detected particles (the two electrons and the $f \! \bar f$ pair), the CM energy
and the  $\tilde \chi_1^0$ and $\tilde \chi_2^0$ masses, which we assume known
from other experiments \cite{tdr,martyn}. In general, it is necessary to have as
many kinematical relations as unknown variables in order to determine the
momenta of the undetected particles. In our case, there are 8 unknowns (the 4
components of the two $\tilde \chi_1^0$ momenta) and 8 constraints. These are
derived from energy and momentum conservation (4 constraints), from the fact
that the two $\tilde \chi_1^0$ are on shell (two constraints), from the decay of
the $\tilde \chi_2^0$ (one constraint) and from the additional hypothesis that
in $e^- e^-$ scattering two particles of equal mass are produced (one
constraint)
\footnote{This
applies for $\tilde e_L \tilde e_L$ and $\tilde e_R \tilde e_R$
production, but not for $\tilde e_R \tilde e_L$, in which case the selectron
masses cannot be reconstructed.}.
These 8 equations determine the 4-momenta of the two $\tilde \chi_1^0$ up to a
4-fold ambiguity, which is partially reduced requiring that
the solutions have positive $p_{\tilde e}^2$ and
that the ``reconstructed'' $m_{\tilde \chi_2^0}^\mathrm{rec}$ is
similar to the real value
\footnote{Although $m_{\tilde \chi_2^0}$ is used as an input for
the reconstruction, the resulting $m_{\tilde \chi_2^0}^\mathrm{rec}$ may be
slightly different from this value, see the paragraph below.}.
If none of the four solutions passes these conditions the event is discarded,
otherwise from the remaining solutions we select the one giving the smallest
$m_{\tilde e}$. For
$\mu^+ \mu^-$ final states this is the best choice: for the events where
the two selectrons are nearly on-shell (these events give the main contribution
to the cross section) the solution with smallest $m_{\tilde e}$ is the
``correct'' one 65\% of the time, and provides a very good reconstruction of the
selectron and $\tilde \chi_2^0$ rest frames. The (discarded) solution with
largest $m_{\tilde e}$ gives a bad $\tilde \chi_2^0$ 4-momentum, and leads to
large distortions for the angular distributions
in the $\tilde \chi_2^0$ rest frame. For $q \bar q$ final states the solution
with smallest $m_{\tilde e}$ gives the best reconstruction as well, but the
difference with the other solution is not so important.

It is worthwhile remarking here that ISR, beamstrahlung, particle width effects
and detector resolution degrade the determination of the $\tilde \chi_1^0$
momenta. However, the reconstruction is successfully achieved in most cases.
The reconstruction procedure determines the momenta of the two unobserved
$\tilde \chi_1^0$,  identifying which selectron has decayed to $e^- \tilde
\chi_1^0$ and which to $e^- \tilde \chi_2^0$, and also distinguishes
between the electrons resulting from each of these decays. The
knowledge of all
the final state momenta, as well as the identification of the particles
resulting from each decay, allows to construct various mass, angular and energy
distributions. These are discussed in turn.

\subsection{Mass distributions}

\begin{figure}[htb]
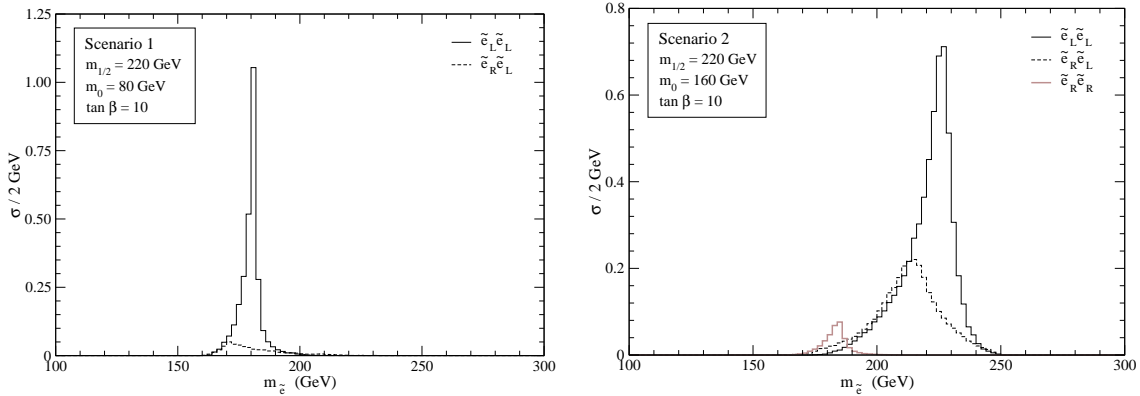

\begin{center}
\begin{tabular}{cc}
\epsfig{file=Figs/rec1.eps,width=7.3cm,clip=} &
\epsfig{file=Figs/rec2.eps,width=7.3cm,clip=}
\end{tabular}
\caption{Reconstructed selectron mass in scenarios 1 and 2, for unpolarised
beams.}
\label{fig:rec}
\end{center}
\end{figure}

Let us call ``$e^-_1$'' and ``$\tilde \chi_{1,1}^0$'' the particles resulting
from $\tilde e \to e^- \tilde \chi_1^0$, with ``$\tilde e_1$'' the corresponding
selectron, and analogously ``$e^-_2$'', ``$\tilde \chi_{1,2}^0$'' and
``$\tilde e_2$'' the particles involved in $\tilde e \to e^- \tilde \chi_2^0$.
The reconstructed mass of the selectrons is simply
\begin{equation}
m_{\tilde e} \equiv \sqrt{p_{\tilde e_1}^2} = 
\sqrt{p_{\tilde e_2}^2} \,,
\end{equation}
with
\begin{eqnarray}
p_{\tilde e_1} & \equiv & p_{e_1^-} + p_{\tilde \chi_{1,1}^0} \,, \nonumber \\
p_{\tilde e_2} & \equiv & p_{e_2^-} + p_{\tilde \chi_{1,2}^0} + p_f + p_{\bar f}
\end{eqnarray}
in obvious notation. All these momenta are taken in the laboratory frame.
The distribution of this variable for the $\tilde e_L \tilde e_L$ and 
$\tilde e_R \tilde e_R$ signals and the
$\tilde e_R \tilde e_L$ background is shown in Fig.~\ref{fig:rec}. In these and
the rest of plots we represent cross sections; the number of observed events
depends on the luminosity and is subject to statistical fluctuations. For the
generation of the distributions we have taken sufficiently high statistics so as
to have a small Monte Carlo uncertainty.

As seen from Fig.~\ref{fig:rec}, the reconstruction of the masses is quite
effective. For $\tilde e_L \tilde e_L$ production, a peak around the true
$\tilde e_L$ mass ($m_{e_L} = 181$ GeV in scenario 1, $m_{e_L} = 227$ GeV in
scenario 2)
is observed in each scenario. (The peak is sharper in scenario 1 due to the
better energy resolution for muons than for jets and the smaller $\tilde e_L$
width.) For $\tilde e_R \tilde e_R$ production in scenario 2, a tiny peak is
observed around the true $\tilde e_R$ mass $m_{e_R} = 185$ GeV.
The $\tilde e_R \tilde e_L$ background is suppressed by the reconstruction
procedure in both scenarios. In scenario 1, its distribution is approximately
flat, but in scenario 2 it noticeably concentrates around $215$
GeV. This behaviour is a result of the smaller ratio
$(m_{\tilde e_L}-m_{\tilde e_R})/(m_{\tilde e_L}+m_{\tilde e_R})$: in scenario
2, the hypothesis of two particles produced with equal mass, used for the
reconstruction, becomes more accurate. The cross sections of the three
processes, before and after the reconstruction, are collected in
Table~\ref{tab:cross}. We also include the cross sections for polarised beams,
in order to show how the $\tilde e_L \tilde e_L$ and $\tilde e_R \tilde e_R$
signals can be enhanced with negative and positive beam polarisation,
respectively, while the $\tilde e_R \tilde e_L$ background is reduced in both
cases.

\begin{table}[htb]
\begin{center}
\begin{small}
\begin{tabular}{ccccccccccccc}
& \multicolumn{4}{c}{Scenario 1} & 
  & \multicolumn{6}{c}{Scenario 2} \\
& \multicolumn{2}{c}{$P_{00}$} & \multicolumn{2}{c}{$P_{--}$} & ~~
  & \multicolumn{2}{c}{$P_{00}$} & \multicolumn{2}{c}{$P_{--}$}
  & \multicolumn{2}{c}{$P_{++}$} \\
& before & after & before & after &
  & before & after & before & after & before & after \\
\hline
$\tilde e_L \tilde e_L$ & 3.25 & 2.88 & 10.52 & 9.32 &
  & 6.52 & 5.71 & 21.14 & 18.50 & 0.26 & 0.23 \\
$\tilde e_R \tilde e_R$ & -- & -- & -- & -- &
  & 0.41 & 0.34 & 0.02 & 0.01 & 1.34 & 1.10 \\
$\tilde e_R \tilde e_L$ & 1.66 & 0.47 & 0.60 & 0.17 &
  & 5.41 & 2.63 & 1.95 & 0.95 & 1.95 & 0.95
\end{tabular}
\caption{Cross sections (in fb) for selectron pair production and subsequent
decay in scenarios 1 and 2. We quote results before and after reconstruction,
for unpolarised beams ($P_{00}$), for $P_1 = P_2 = -0.8$ ($P_{--}$) and in
scenario 2 also for $P_1 = P_2 = 0.8$ ($P_{++}$). Detector cuts are included in
all cases.}
\label{tab:cross}
\end{small}
\end{center}
\end{table}

The experimental measurement of the selectron masses from these distributions is
less precise than from threshold scans \cite{feng:2001,blochinger}.
These mass distributions are however useful in order to identify
$\tilde e_L \tilde e_L$ and $\tilde e_R \tilde e_R$ production, and may be
used to separate them from the $\tilde e_R \tilde e_L$ background.
We also note that in the $e^- \tilde \chi_1^0 e^- \tilde \chi_1^0$ channel,
even without a complete determination of the final state momenta
the minimum kinematically allowed selectron mass can be obtained
\cite{feng1993}. The kinematical distribution of this quantity also peaks at
the true selectron masses.

\subsection{Production angle}

\begin{figure}[htb]
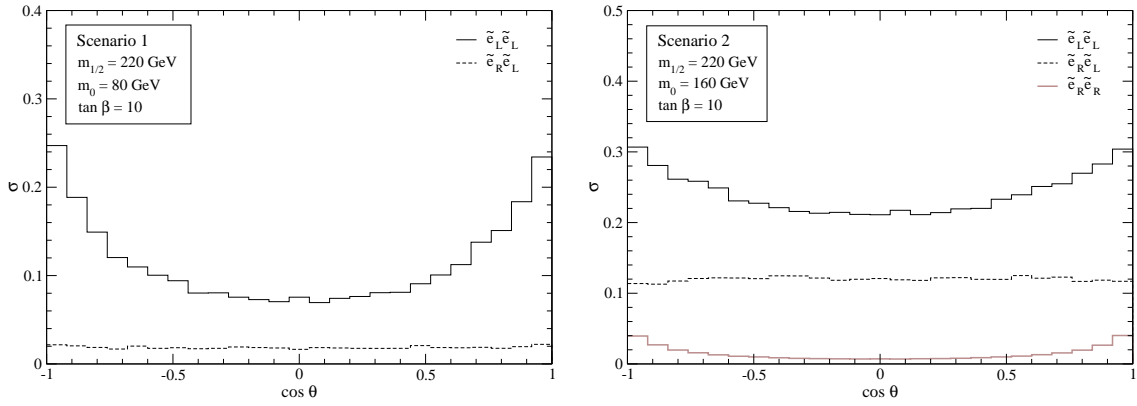

\begin{center}
\begin{tabular}{cc}
\epsfig{file=Figs/rec1-thprod.eps,width=7.3cm,clip=} &
\epsfig{file=Figs/rec2-thprod.eps,width=7.3cm,clip=}
\end{tabular}
\caption{Dependence of the cross section on the production angle $\theta$ in
scenarios 1 and 2, for unpolarised beams.}
\label{fig:rec-thprod}
\end{center}
\end{figure}

The determination of the selectron momenta allows the study of the
dependence of the cross section on the production angle $\theta$. In $e^+ e^-$
scattering, this analysis can be used to determine that the selectrons
are spinless particles \cite{tdr}
\footnote{In the $e^- \tilde \chi_1^0 e^-
\tilde \chi_1^0$ channel, to determine the final state momenta the conditions
$p_{\tilde e}^2 = m_{\tilde e}^2$ are required from the beginning.}.
In our case, it does not provide information about the
selectron spins. This distribution, shown in Fig.~\ref{fig:rec-thprod} for
both scenarios, is symmetric
with respect to the value $\cos \theta = 0$, as must be for a process with a
symmetric initial state $e^- e^-$, and follows the expected shape.

\subsection{Electron angular distributions in selectron rest frames}

\begin{figure}[htb]
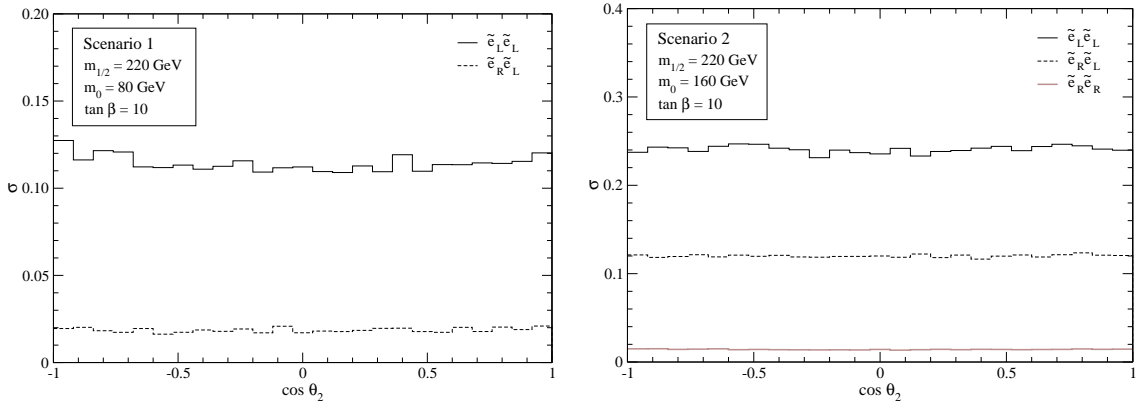

\begin{center}
\begin{tabular}{cc}
\epsfig{file=Figs/rec1-th2.eps,width=7.3cm,clip=} &
\epsfig{file=Figs/rec2-th2.eps,width=7.3cm,clip=}
\end{tabular}
\caption{Angular distribution of $e_2^-$ with respect to the positive $y$ axis
in the $\tilde e_2$ rest frame, in scenarios 1 and 2, for unpolarised beams.}
\label{fig:rec-th2}
\end{center}
\end{figure}

The selectron spins can be effectively analised through the angular
distributions of the electrons in the selectron rest frames. Since the
selectrons
are spinless particles, their decay is isotropic. Therefore in each selectron
rest frame the angular distribution of the produced electron (and neutralino)
with respect to any fixed direction must be flat. In Fig.~\ref{fig:rec-th1}
we show the dependence of the cross section on the angle $\theta_1$
between the momentum of $e_1^-$ in the $\tilde e_1$ rest frame
and the positive $y$ axis (remember that $e_1^-$ and $\tilde e_1$ correspond
to the decay
$\tilde e \to e^- \tilde \chi_1^0$, and are identified in the reconstruction
process). The analogous is shown for the decay $\tilde e \to e^- \tilde
\chi_2^0$ in Fig.~\ref{fig:rec-th2}, with $\theta_2$ the angle between the
momentum of $e_2^-$ and the positive $y$ axis.
Similar distributions can be obtained for
the $x$ and $z$ axes, proving that the decay of both selectrons is isotropic.
 
\begin{figure}[htb]
\begin{center}
\begin{tabular}{cc}
\epsfig{file=Figs/rec1-th1.eps,width=7.3cm,clip=} &
\epsfig{file=Figs/rec2-th1.eps,width=7.3cm,clip=}
\end{tabular}
\caption{Angular distribution of $e_1^-$ with respect to the positive $y$ axis
in the $\tilde e_1$ rest frame, in scenarios 1 and 2, for unpolarised beams.}
\label{fig:rec-th1}
\end{center}
\end{figure}

\subsection{Electron energy distributions in laboratory frame}

\begin{figure}[htb]
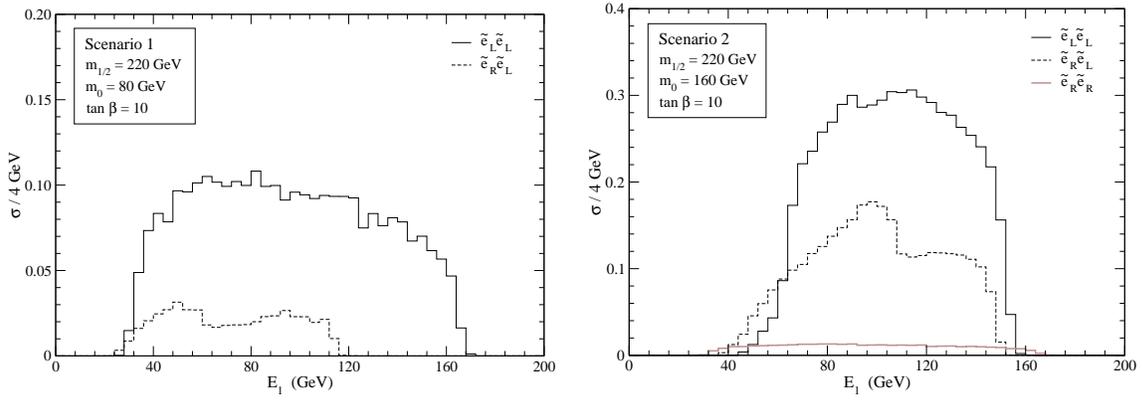

\begin{center}
\begin{tabular}{cc}
\epsfig{file=Figs/rec1-E1.eps,width=7.3cm,clip=} &
\epsfig{file=Figs/rec2-E1.eps,width=7.3cm,clip=}
\end{tabular}
\caption{Energy distribution of the electron resulting from $\tilde e \to e^-
\tilde \chi_1^0$, in scenarios 1 and 2, for unpolarised beams.}
\label{fig:rec-E1}
\end{center}
\end{figure}

The masses and spins of the selectrons can be further investigated via the
electron energy distributions. 
In the selectron rest frame, the electron energy is fixed by the kinematics of
the 2-body decay, and furthermore their decay is isotropic. 
Then, for $\tilde e_L \tilde e_L$ and $\tilde e_R \tilde e_R$
production the electron energy distributions are
flat, with end points at
\begin{eqnarray}
E_i^\mathrm{max} & = & \frac{\sqrt s}{4}
\left( 1- \frac{m_{\tilde \chi_i^0}^2}{m_{\tilde e_{L,R}}^2} \right) (1 + \beta)
\;, \nonumber \\
E_i^\mathrm{min} & = & \frac{\sqrt s}{4}
\left( 1- \frac{m_{\tilde \chi_i^0}^2}{m_{\tilde e_{L,R}}^2} \right) (1 - \beta)
\;,
\label{ec:end}
\end{eqnarray}
where $\beta = \sqrt{1-4 m_{\tilde e_{L,R}}^2/s}$. For mixed selectron
production the electron energy spectra are flat as well, but the expressions of
the end points are more involved. In contrast with the decay mode
$e^- \tilde \chi_1^0 e^- \tilde \chi_1^0$, in the channel
$e^- \tilde \chi_1^0 e^- \tilde \chi_2^0$ there are two different distributions
for the energies $E_1$ and $E_2$ of the electrons $e_1^-$ and $e_2^-$,
respectively. The distribution of $E_1$ in both scenarios is shown in
Fig.~\ref{fig:rec-E1}. From Eqs.~\ref{ec:end}, in scenario 1 the expected end
points are $E_1^\mathrm{min} = 30$ GeV, $E_1^\mathrm{max} = 166$ GeV.
In scenario 2, for $\tilde e_L$ decays the expected limits of the distributions
are $E_1^\mathrm{min} = 63$ GeV, $E_1^\mathrm{max} = 153$ GeV,
and for $\tilde e_R$ decays they are
$E_1^\mathrm{min} = 32$ GeV, $E_1^\mathrm{max} = 166$ GeV. Although the
distributions are smeared by ISR, beamstrahlung and detector effects, 
all these end points
can be clearly observed in the plots. 
However, in the real experiment the different contributions will be summed, and
except in some cases where the end points coincide, in general beam polarisation
will be essential in order to enhance one of the signals and reduce the $\tilde
e_R \tilde e_L$ background. The polarisation also improves the statistics and
thus the acuracy of the end point determination. The measurement of these
quantities gives further evidence that the
selectrons are spinless particles and provides independent determinations of
their masses.

\begin{figure}[htb]
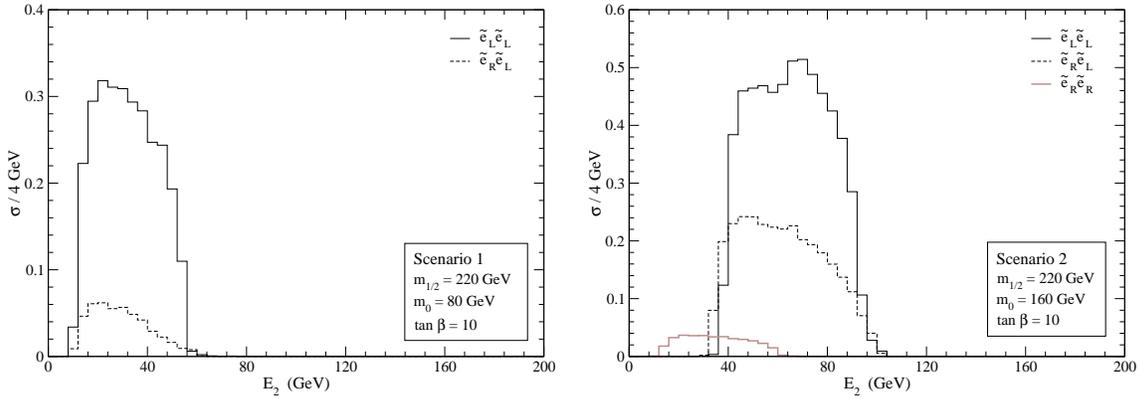

\begin{center}
\begin{tabular}{cc}
\epsfig{file=Figs/rec1-E2.eps,width=7.3cm,clip=} &
\epsfig{file=Figs/rec2-E2.eps,width=7.3cm,clip=}
\end{tabular}
\caption{Energy distribution of the electron resulting from $\tilde e \to e^-
\tilde \chi_2^0$, in scenarios 1 and 2, for unpolarised beams.}
\label{fig:rec-E2}
\end{center}
\end{figure}

For the decays $\tilde e \to e^- \tilde \chi_2^0$, the corresponding electron
energy distribution 
is shown in Fig.~\ref{fig:rec-E2}. In scenario 1, the expected limits are
$E_2^\mathrm{min} = 10$ GeV, $E_2^\mathrm{max} = 55$
GeV. In scenario 2, for $\tilde e_L$ decays the expected end points 
are $E_2^\mathrm{min} = 38$~GeV, $E_2^\mathrm{max} = 93$ GeV
and for $\tilde e_R$ decays they are $E_2^\mathrm{min} = 12$ GeV,
$E_2^\mathrm{max} = 60$ GeV. All these end points can be observed in
Fig.~\ref{fig:rec-E2}.

\subsection{Distributions in $\boldsymbol{\tilde \chi_2^0}$ rest frame}

\begin{figure}[htb]
\begin{center}
\begin{tabular}{ccc}
\epsfig{file=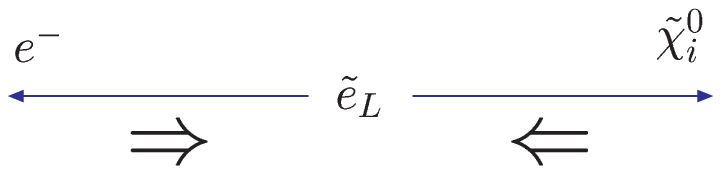,width=5cm,clip=} & ~~~~ & 
\epsfig{file=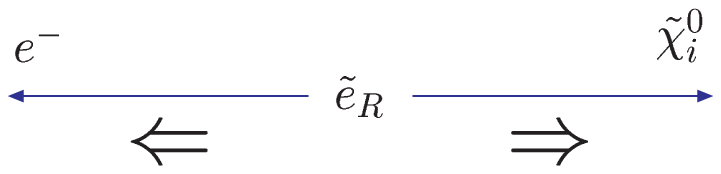,width=5cm,clip=} \\
(a) & & (b)
\end{tabular}
\caption{Helicity of the neutralinos produced in $\tilde e_L$ and $\tilde e_R$
decays.}
\label{fig:Xhel}
\end{center}
\end{figure}

The neutralinos resulting from selectron decay are 100\% polarised in the
direction of their momentum if selectron mixing is neglected. This fact can be
explained as follows. The term of
the Lagrangian describing $e \tilde e \tilde \chi_i^0$ interactions is,
neglecting selectron mixing,
\begin{equation}
\mathcal{L}_{e \tilde e \tilde \chi_i^0} =
a_L^i \; \tilde e_L^* \, \bar{\tilde \chi}_i^0  P_L \, e
+ a_L^{i*} \; \tilde e_L \, \bar e \, P_R \, \tilde \chi_i^0
+ a^{i}_R \; \tilde e_R^* \, \bar{\tilde \chi}_j^0  P_R \, e
+ a_R^{i*} \; \tilde e_R \, \bar e \, P_L \, \tilde \chi_i^0 \,,
\end{equation}
with $a_L^i$ and $a_R^i$ constants.
For $\tilde e_L \to e^- \tilde \chi_i^0$, the produced electron has chirality
$-1$, and the neutralino has chirality $+1$. For a massive
particle, the helicity eigenstates do not coincide in general with the chirality
eigenstates.
However, since in this case the helicity of the electron is $-1$ and the
selectron is spinless, angular momentum conservation implies that the neutralino
must have negative helicity, as shown
schematically in Fig.~\ref{fig:Xhel}a. For $\tilde e_R$ decays, the same
argument shows that the produced neutralinos have positive helicity
(Fig.~\ref{fig:Xhel}b)

\begin{figure}[htb]
\begin{center}
\epsfig{file=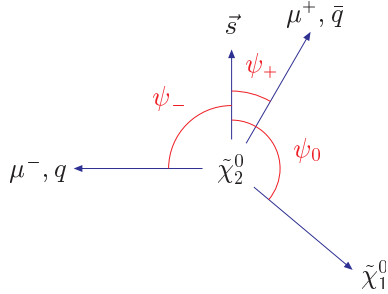,width=5cm,clip=}
\caption{Definition of the polar angles $\psi_+$, $\psi_-$ and $\psi_0$.}
\label{fig:X2dist}
\end{center}
\end{figure}

The expressions of the polarised differential decay widths of
$\tilde \chi_2^0 \to \tilde \chi_1^0 f \! \bar f$ are rather involved
\cite{X2decay1,X2decay2,X2decay3}. However, the angular distribution of a single
decay product in the $\tilde \chi_2^0$ rest frame can be cast in a very compact
form. Let us define $\psi_+$, $\psi_-$ and $\psi_0$ as the polar angles between
the 3-momenta in the $\tilde \chi_2^0$ rest frame
of $\bar f=\mu^+,\bar q$, $f=\mu^-,q$ and $\tilde \chi_1^0$, respectively, and
the $\tilde \chi_2^0$ spin $\vec s$. These angles are represented in
Fig.~\ref{fig:X2dist}.

Integrating all the variables except $\psi_+$, $\psi_-$ or $\psi_0$, the
angular decay distributions can be written as
\begin{eqnarray}
\frac{1}{\sigma} \frac{d\sigma}{d\cos \psi_+} & = & \frac{1+ h^f_+ \cos \psi_+}
{2} \,, \nonumber \\
\frac{1}{\sigma} \frac{d\sigma}{d\cos \psi_-} & = & \frac{1+ h^f_- \cos \psi_-}
{2} \,, \nonumber \\
\frac{1}{\sigma} \frac{d\sigma}{d\cos \psi_0} & = & \frac{1+ h^f_0 \cos \psi_0}
{2} \,.
\end{eqnarray}
The constants $h_i^f$ depend on the type of fermion $f=\mu,u,d,s,c,b$ considered
(and of course on the SUSY scenario),
and measure the degree of correlation between the $\tilde \chi_2^0$ spin and the
direction in which the fermion $f$ is emitted.
If CP is conserved in this decay, the Majorana nature of
$\tilde \chi_1^0$ and $\tilde \chi_2^0$
implies that $h_+^f=-h_-^f$ and $h_0^f=0$. In scenario 1,
$h_+^\mu = -h_-^\mu = -0.57$, and in scenario 2 we find
$h_+^b = -h_-^b = 0.75$.

The fact that the $\tilde \chi_2^0$ produced in $\tilde e$ decays are polarised
allows the study of angular distributions in the $\tilde \chi_2^0$ rest frame.
In analogy with Fig.~\ref{fig:X2dist}, we define the polar angles  $\varphi_+$,
$\varphi_-$ and $\varphi_0$ between the 3-momenta in the $\tilde \chi_2^0$ rest
frame of $\bar f$, $f$ and $\tilde \chi_1^0$, respectively, and the
3-momentum $\vec p$ of $\tilde \chi_2^0$ in the $\tilde e_2$ rest frame. (In
$\tilde e_R$ decays, $\vec p$ gives the direction of the spin $\tilde s$, and
in $\tilde e_L$ decays the opposite direction.) With these definitions, we build
the spin asymmetries
\begin{equation}
A_i \equiv \frac{N(\cos \varphi_i > 0) - N(\cos \varphi_i < 0)}{N(\cos
\varphi_i> 0) + N(\cos \varphi_i < 0)}  \,,
\end{equation}
where $N$ stands for the number of events. The theoretical prediction for these
asymmetries is $A_i = \lambda \, h_i^f/2$, with $\lambda$ the helicity
of $\tilde \chi_2^0$.

\begin{figure}[htb]
\begin{center}
\epsfig{file=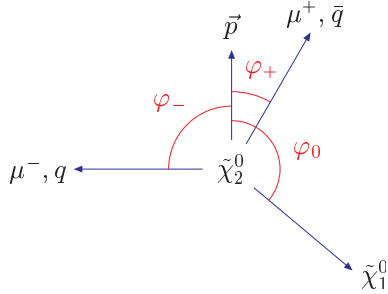,width=5cm,clip=}
\caption{Definition of the polar angles $\varphi_+$, $\varphi_-$ and
$\varphi_0$.}
\label{fig:X2dist2}
\end{center}
\end{figure}

\begin{figure}[htb]
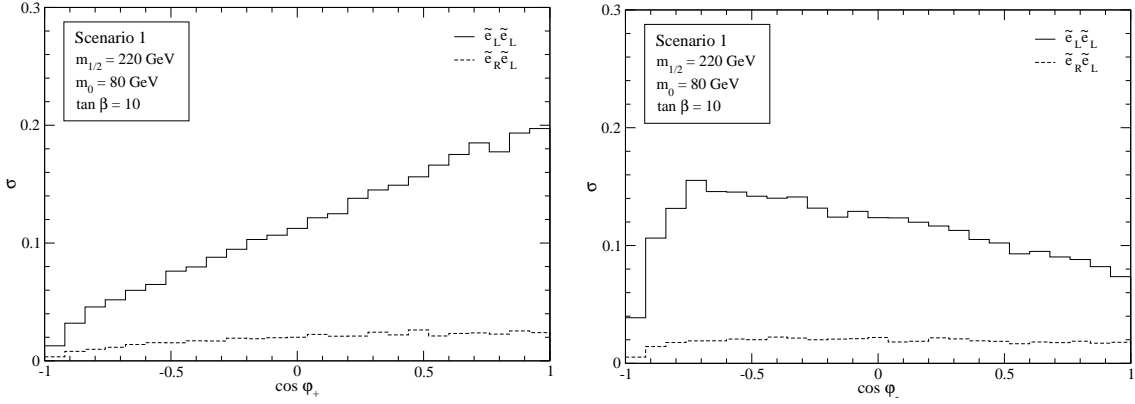

\begin{center}
\begin{tabular}{cc}
\epsfig{file=Figs/rec1-phi+.eps,width=7.3cm,clip=} &
\epsfig{file=Figs/rec1-phi-.eps,width=7.3cm,clip=}
\end{tabular}
\caption{Angular distribution of the $\mu^+$ and $\mu^-$ in the $\tilde
\chi_2^0$ rest frame with respect to the $\tilde \chi_2^0$ momentum in the
$\tilde e_2$ rest frame, in scenario 1.}
\label{fig:rec1-phi}
\end{center}
\end{figure}

The angular distributions of $\mu^+$ and $\mu^-$ in scenario 1 are presented in
Fig.~\ref{fig:rec1-phi}, for the $\tilde e_L \tilde e_L$ signal and the $\tilde
e_R \tilde e_L$ background. In both cases we observe a suppression at $\cos
\varphi_\pm   = -1$, due to the detector cuts. For $\cos \varphi_\pm = -1$, the
$\mu^\pm$ is emitted in the direction of $-\vec p$, which is the direction of
the $e^-_2$, thus these two particles are not isolated and the event is rejected
by the requirement on ``lego-plot'' separation.

The slopes of these distributions clearly show that the $\tilde \chi_2^0$ has
helicity $-1$ ($h_+^\mu = -0.57$ in this scenario) and hence that the decaying
selectron is a $\tilde e_L$. This information can be of course obtained from 
other sources, for instance with the comparison of production cross sections
with and without polarisation. Additionally, these plots demonstrate
that the $\tilde \chi_2^0$ has nonzero spin (compare with $\tilde e$ decays in
Figs.~\ref{fig:rec-th1} and
\ref{fig:rec-th2}). For $\tilde e_L \tilde e_L$ production only
\footnote{The experimentally measured asymmetries would also include
the contribution from $\tilde e_R \tilde e_L$ production, which might be
reduced using beam polarisation. For simplicity we quote the results
for $\tilde e_L \tilde e_L$ exclusively.},
the asymmetries are $A_+ = 0.39$, $A_- = -0.12$, while the theoretical
expectations are $A_\pm = \pm 0.29$.

\begin{figure}[htb]
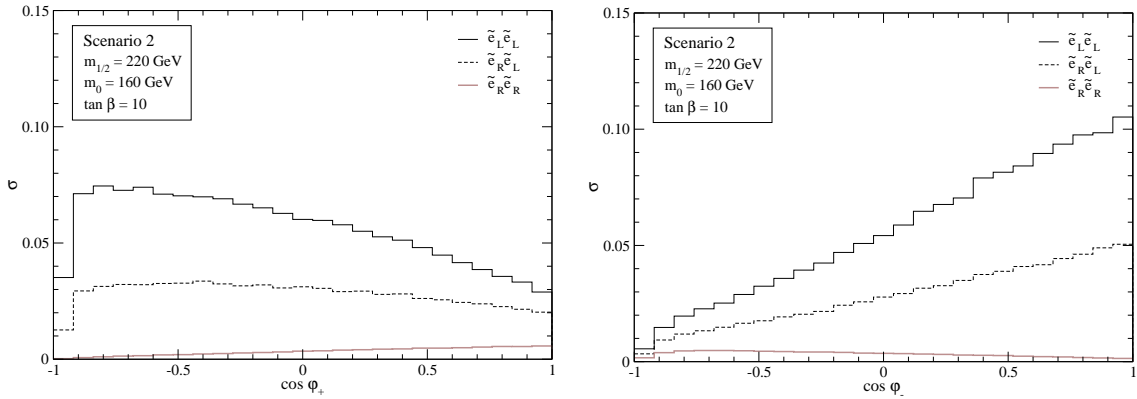

\begin{center}
\begin{tabular}{cc}
\epsfig{file=Figs/rec2-phi+.eps,width=7.3cm,clip=} &
\epsfig{file=Figs/rec2-phi-.eps,width=7.3cm,clip=}
\end{tabular}
\caption{Angular distribution of the $\bar b$ ($\varphi_+$) and $b$
($\varphi_-$)in the $\tilde
\chi_2^0$ rest frame with respect to the $\tilde \chi_2^0$ momentum in the
$\tilde e_2$ rest frame, in scenario 2.}
\label{fig:rec2-phi}
\end{center}
\end{figure}

In scenario 2, the experimental measurement of these asymmetries requires to
distinguish $q$ from $\bar q$. This is very difficult to do in general, so we
restrict ourselves to $q=b$ where this is possible although with a limited
efficiency. The angular
distributions of the $\bar b$
($\varphi_+$) and $b$ ($\varphi_-$) are shown in Fig.~\ref{fig:rec2-phi}.
In these plots we have not taken into account neither efficiencies nor
mistagging rates for the $b,\bar b$ identification.

The slope of the angular distribution for the $\tilde e_L \tilde e_L$ signal
again shows that
the $\tilde \chi_2^0$ has negative helicity ($h_+^b = 0.75$ in this scenario),
and thus indicates that the decaying selectron is a $\tilde e_L$. The
asymmetries for this process alone are $A_+ = -0.15$, $A_- = 0.41$, and the
theoretical predictions $A_\pm = \mp 0.37$. For $\tilde e_R \tilde e_R$, the
positive slope indicates a decay from a $\tilde e_R$, and the asymmetries in
this case are $A_+ = 0.39$, $A_- = -0.21$.
The $\tilde e_R \tilde e_R$ signal could in principle be observable with
positive beam polarisation, which increases its cross section by a
factor of 3.24 and reduces
$\tilde e_L \tilde e_L$ and $\tilde e_R \tilde e_L$ by factors of 0.04 and 0.36,
respectively (see Table~\ref{tab:cross}). In addition, kinematical cuts on
reconstructed masses could be applied to enhance $\tilde e_R \tilde e_R$
(see Fig.~\ref{fig:rec}).

\section{Conclusions}
\label{sec:4}

In this note we have analysed $\tilde e_L \tilde e_L$ and
$\tilde e_R \tilde e_R$
production in $e^- e^-$ collisions, with subsequent decay
$\tilde e \tilde e \to e^- \tilde \chi_1^0 e^- \tilde \chi_2^0
\to e^- \tilde \chi_1^0 e^- \tilde \chi_1^0 f \! \bar f$.
For $\tilde e_R \tilde e_R$ production, this is a rare channel, with a cross
section much smaller than the leading mode
$e^- \tilde \chi_1^0 e^- \tilde \chi_1^0$, but for $\tilde e_L \tilde e_L$ it
may have a comparable or even larger cross section. We have 
shown some of the benefits that the reconstruction of all
the final state momenta offers, allowing the study of mass, angular and 
energy distributions. 

We have demonstrated that in this decay mode it is possible to gather
information on the $\tilde
\chi_2^0$ spin, what is not possible in $e^- \tilde \chi_1^0 e^- \tilde
\chi_1^0$ decays. This is of special interest since in selectron decays 
the neutralinos are 100\% polarised, having negative helicity (in the selectron
rest frame) in $\tilde e_L$ decays and positive helicity in $\tilde e_R$ decays.
Indeed, $\tilde e_L$ decays are a source of 100\% polarised $\tilde \chi_2^0$
with a cross section comparable or even larger than direct production
$e^+ e^- \to \tilde \chi_1^0 \tilde \chi_2^0$. In this work we have used
the distribution of the $\tilde \chi_2^0$ decay products only to distinguish
$\tilde e_L$ from $\tilde e_R$, but $\tilde \chi_2^0$ decays also offer a good
place to investigate CP violation in the neutralino sector \cite{CP} through
the analysis of CP-violating asymmetries \cite{triple}. 
This study will be presented elsewhere \cite{next}.

\vspace{1cm}
\noindent
{\Large \bf Acknowledgements}

\vspace{0.4cm} \noindent
I thank A. M. Teixeira for previous collaboration. This work has been supported
by the European Community's Human Potential Programme under contract
HTRN--CT--2000--00149 Physics at Colliders and by FCT
through project CFIF--Plurianual (2/91) and grant SFRH/BPD/12603/2003.


\begin{thebibliography}{99}
\bibitem{tdr}
J. A. Aguilar-Saavedra {\it et al.}  [ECFA/DESY LC Physics Working Group
Collaboration], hep-ph/0106315

\bibitem{old}
J. A. Aguilar-Saavedra and A. M. Teixeira, Nucl. Phys. {\bf B} 675, 70 (2003)
[hep-ph/0307001]

\bibitem{martyn}
H. U. Martyn and G. A. Blair, hep-ph/9910416

\bibitem{feng:2001}
J. L. Feng and M. E. Peskin, Phys. Rev. D {\bf 64}, 115002 (2001)
[hep-ph/0105100]

\bibitem{blochinger}
C. Blochinger, H. Fraas, G. Moortgat-Pick and W. Porod, Eur. Phys. J. C {\bf
24}, 297 (2002) [hep-ph/0201282]

\bibitem{feng1993}
J. L. Feng and D. E. Finnell, Phys. Rev. D {\bf 49}, 2369 (1994) 
[hep-ph/9310211]

\bibitem{martyn2}
H. U. Martyn, hep-ph/0002290

\bibitem{X2decay1}
G. Moortgat-Pick, H. Fraas, A. Bartl and W. Majerotto, Eur. Phys. J. C {\bf 9},
521 (1999) [Erratum-ibid.\ C {\bf 9}, 549 (1999)] [hep-ph/9903220]
\bibitem{X2decay2}
S. Y. Choi, H. S. Song and W. Y. Song, Phys. Rev. D {\bf 61}, 075004 (2000)
[hep-ph/9907474]
\bibitem{X2decay3}
A. Djouadi, Y. Mambrini and M. M\"uhlleitner, Eur. Phys. J. C {\bf 20}, 563
(2001) [hep-ph/0104115]

\bibitem{CP}
See for instance J. Kalinowski,
Acta Phys. Polon. B {\bf 34}, 3441 (2003)
[hep-ph/0306272].
For the reconstruction of CP-violating quantities from CP-conserving
observables, see 
S. Y. Choi, J. Kalinowski, G. Moortgat-Pick and P. M. Zerwas,
Eur. Phys. J. C {\bf 22}, 563 (2001)
[Addendum-ibid.\ C {\bf 23}, 769 (2002)]
[hep-ph/0108117]

\bibitem{triple}
A. Bartl, H. Fraas, O. Kittel and W. Majerotto, hep-ph/0308141

\bibitem{next}
J. A. Aguilar-Saavedra, hep-ph/0403243, to be published in Phys. Lett. B

\end{thebibliography}
\end{document}